\documentclass[aps,prb,reprint,floatfix]{revtex4-2}
\usepackage{graphicx,graphics}
\usepackage{epstopdf} 
\usepackage{amsmath,amssymb,amsfonts}
\usepackage{latexsym,verbatim}
\usepackage{bm}
\usepackage{bbold}
\usepackage{xcolor}
\usepackage[normalem]{ulem}
\usepackage[breaklinks=false,colorlinks,citecolor=blue,linkcolor=blue,urlcolor=blue]{hyperref}
\usepackage{textcomp} 
\usepackage{mathtools}
\usepackage{braket}
\usepackage{soul}    
\usepackage{multirow}

\newif\ifdraft
\drafttrue 

\definecolor{Mygrey}{gray}{0.80}
\definecolor{lteal}{rgb}{0.10,0.60,0.70}
\definecolor{dkred}{rgb}{0.80,0.10,0.00}
\definecolor{dgreen}{rgb}{0.0, 0.5, 0.0}

\ifdefined\LightComments
  \newcommand\SP[2]{\textcolor{orange}{#2}}

\else
  \newcommand\SP[2]{\textcolor{orange}{\begingroup\csname @fileswfalse\endcsname\sout{#1}\endgroup{~#2}}}

 \fi

\begin{document}

\title{Genuine multipartite entanglement in ring-shaped Hubbard models}

\author{Khuram Shahzad$^1$}
\author{Stefano Pittalis$^2$}
\author{Guido Goldoni$^1$}
\author{Rosa Di Felice$^{2,3}$}
\author{Filippo Troiani$^2$}
\email{filippo.troiani@nano.cnr.it}

\affiliation{$^1$Dipartimento di Scienze Fisiche, Informatiche e Matematiche, Università di Modena e Reggio Emilia, Modena, Italy}
\affiliation{$^2$Istituto Nanoscienze -- CNR, S3, Via G. Campi 213/A, 41125 Modena, Italy}
\affiliation{$^3$Department of Physics and Astronomy and Department of Quantitative and Computational Biology, University of Southern California, Los Angeles, CA 90089}

\begin{abstract}
Spin states can be extracted from many-electron systems by detecting electrons at a set of lattice sites, which act as particle labels. Here we quantify the genuine multipartite entanglement (GME) of such spin states, which are generally mixed, in standard and extended Hubbard rings, together with the corresponding detection probability. We show that the on-site repulsion, by suppressing charge fluctuations, enhances the GME, and that in the strongly interacting regime at half filling the detection probability decreases only weakly with the number of spins. The GME is robust at temperatures below the singlet--triplet gap, and can even be enhanced by doping, in spite of the induced charge fluctuations. The transition from spin- to charge-density waves in the extended Hubbard model is signaled by an abrupt suppression of both the GME and the detection probability.
\end{abstract}

\date{\today}

\maketitle

\section{Introduction}

The theory of entanglement between distinguishable particles (spins, qubits) has been extensively developed, also in view of its relevance for the quantum technologies \cite{Horodecki09a}. Unlike qubits, bosons and fermions cannot be individually addressed. Correspondingly, their states are defined in Hilbert spaces that are not naturally endowed with a tensor-product structure. This makes the definition and quantification of entanglement more ambiguous and problematic~\cite{Amico08a,Benatti20a,Sun22a}. 

In the case of bosons, indistinguishability allows the generation of entanglement and represents a resource for quantum information processing \cite{Mahdavipour24a,Piccolini_2025}. Entanglement also represents a means to characterize electronic states in atoms, molecules, and materials~\cite{Schliemann01a, Li01a,Zanardi02a,Amico08a,Franca08a,Gigena15a,Franca11a}.
Approaches to fermionic entanglement are mostly based on the concepts of particles or modes \cite{Benatti20a}.
Orbital entanglement \cite{Zanardi02a,Legeza2003,Rissler2006,Boguslawski2015} belongs to the latter class, and has found useful application in the domain of quantum chemistry \cite{Boguslawski2013,Szalay17a,Duperrouzel2015,Stein2016}. Localized orbitals have long been used to interpret electronic structure and derive effective model parameters, including through orthogonal valence-bond approaches~\cite{Angeli24a}. Their role in describing electronic correlations has also been explored through comparisons of mutual information in localized and delocalized orbital bases~\cite{Tenti24} and through the quantification of physical entanglement between localized orbitals~\cite{Ding24a}. The maximum bipartite orbital entanglement attainable from Slater determinants under orbital rotations has also been determined~\cite{AlivertiPiuri26}, complementing studies of genuine multipartite entanglement in extracted spin states~\cite{Troiani25a}. Its operational interpretation requires accounting for restrictions on local operations, including particle-number superselection rules~\cite{Wiseman03a,Ding2021}.

Here we adopt an alternative approach, in which a fermionic state is reduced to a spin state by freezing the spatial degrees of freedom and using a set of spatial modes (orbitals) as labels for the particles, within a detection procedure~\cite{Benatti20a,Killoran14a,Bouvrie17a,LoFranco18a,Troiani25a,Troiani25b}. The properties of the spin state depend both on those of the original fermionic state and on the choice of the detection modes, which can also be identified --- as a limiting case --- with the spatial positions \cite{Vedral03a,Oh04a,Pittalis15a}. Such a dependence can in principle be exploited to tailor the entanglement properties of the spin state but, on the other hand, introduces an undesirable element of arbitrariness at the conceptual level. In the present paper, we adopt a ``natural'' choice for the detection modes, which are identified with the sites of a lattice model. 

Throughout this paper, the terms ``detection'' and ``extraction'' are used in a formal sense: they refer to the action of annihilation operators on the fermionic state. The detected spin state is thus given by a set of Green's functions, which define an $n$-body reduced density matrix ($n$BRDM). Identifying concrete procedures for extracting particles from a fermionic system, though relevant to the experimental and technological implications of the approach, is beyond the scope of the present work.

We apply this approach to eigenstates and thermal-equilibrium states of standard and extended Hubbard models \cite{Hubbard2013}. The investigations on entanglement in the standard Hubbard model have mainly focused on bipartite entanglement, and more specifically entanglement between pairs of spins or between complementary blocks, each one consisting in a set of neighboring sites. These forms of bipartite entanglement were shown to capture quantum phase transitions, or precursors thereof, taking place in finite-size systems. The extended Hubbard model differs from the standard one for the presence of nearest-neighbor interactions. In real systems, these can originate from Coulomb or dipolar forces, and correspondingly take a repulsion or an attractive character. 
At half filling, the physics is governed by the competition between on-site repulsion, which tends to localize one electron per site, and the repulsion between neighboring sites, which penalizes the simultaneous occupation of neighboring sites. The resulting charge fluctuations have a strong and detrimental impact both on the multipartite spin entanglement and on the detection probability.

In a recent Letter~\cite{Troiani25a}, it was demonstrated that, contrary to common wisdom~\cite{Lunkes05a}, GME can be abundantly extracted from single Slater determinants, and maximized via spin-independent transformations derived from the quantum Fourier transform. Close to maximal GME was also shown to result from correlated eigenstates of prominent interacting electron models, ranging from Hubbard to atomistic Hamiltonians, through physically motivated localized orbitals. In Ref.~\cite{Troiani25b}, the authors derived the structure of the spin states extracted from closed-shell states, and identified double occupations and rotational invariance as the factors that limit the extractable entanglement. In these works, the GME was quantified only for pure states of $n$ spins, extracted from pure states of $K$ electrons, with $n=K$. Here we lift both restrictions. For $n<K$, the extracted spin states are mixed, and their GME is quantified via the convex-roof extension of the GME concurrence. Besides, we consider mixed fermionic states, namely thermal-equilibrium states and the degenerate ground multiplets of doped rings. This brings out the role of charge fluctuations, which do not affect the GME for $n=K$, and allows us to address the effects of temperature, doping, and nearest-neighbor interactions. We finally compare the results for the Hubbard ring with those obtained for an atomistic model of the the benzene molecule.

The rest of the paper is organized as follows. In Sec. \ref{sec:method} we illustrate the theoretical and numerical approach utilized for quantifying multipartite entanglement. Section \ref{sec:results} is devoted to the numerical results. The conclusions are drawn in Sec. \ref{sec:conclusions}.

\section{Method\label{sec:method}}

The method consists in the following steps. First we compute the fermionic state, by diagonalizing the relevant Hamiltonian. Then, we derive the state of the extracted spins, and quantify its entanglement through the GME concurrence (Subsec. \ref{subsecA}). For mixed spin states, this calculation implies a numerical minimization, which is implemented through a conjugate gradient approach (Subsec. \ref{sec:numerics}).

\subsection{Theoretical approach\label{subsecA}} 

As mentioned in the previous Section, the $K$-fermion system can be reduced to a system of $n$ spins by means of a detection procedure, where the spatial degrees of freedom are frozen and used for particle labeling \cite{Benatti20a,Killoran14a,Bouvrie17a,LoFranco18a}.
In this ``effective-distinguishability'' approach, the entanglement properties of the extracted $n$-spin state thus result both from the properties of the original $K$-electron state $|\Phi_e\rangle$ and from the chosen set of orbitals (modes) that define the particle extraction. In the case of the Hubbard model, the spins are naturally identified through their spatial localization, and the generic orbital $\phi_k$ corresponds to the $k$-th site of the system.

The $n$-spin state extracted from the orbitals $\phi_1,\dots,\phi_n$ is formally given by Green's functions, and specifically by the $n$-body reduced density matrices ($n$BRDMs) \cite{pilar2001elementary}:
\begin{gather}\label{eqn:nBRDM}
    \Gamma^{(n)}_{{\bf\sigma},{\bf\sigma}'} \!=\! \langle \Phi_e | \,\hat{c}_{n,\sigma_n'}^\dagger \dots \hat{c}_{1,\sigma_1'}^\dagger \hat{c}_{1,\sigma_1} \dots \hat{c}_{n,\sigma_n}\, |\Phi_e\rangle\;.
\end{gather}
Here, $\hat{c}^\dagger_{k,\sigma_k}$ and $\hat{c}_{k,\sigma_k}$ respectively create and annihilate an electron in the orbital state $|\phi_k\rangle$ with spin orientation $\sigma_k=\uparrow,\downarrow$; the symbols $\sigma$ and $\sigma'$ that appear in the subscript of the $n$BRDM represent $n$-dimensional vectors, with components $\sigma_k$. The extracted spin state $\rho_n$ is given by the normalized $n$BRDM: 
\begin{gather}
\rho_{n} = \frac{\Gamma^{(n)}_{{\bf\sigma},{\bf\sigma}'}}{{\rm Tr}\left[\Gamma^{(n)}_{{\bf\sigma},{\bf\sigma}'}\right]}\,\,.
\end{gather}
The quantities
\begin{gather}
w_n = {\rm Tr}\left[\Gamma^{(n)}_{{\bf\sigma},{\bf\sigma}'}\right]\
\end{gather}
and
\begin{gather}
p_n = 2^{-\min\,(n,K-n)}\,w_n 
\end{gather}
represent, respectively, the unnormalized and normalized probabilities of extracting $n$ particles from the fermionic state, one for each of the orbitals $\phi_k$ \footnote{The diagonal of the $n$BRDM $\Gamma^{(n)}_{{\bf\sigma},{\bf\sigma}'}$ has $2^n$ elements, each of which corresponds to the joint occupation probability of a combination of $n$ spin-orbitals, and can take values between 0 and 1. However, the sum of the occupation probabilities cannot exceed the exponential of the detection orbitals that can be doubly occupied ($K-n$). Hence the normalization to $2^{\min(n,K-n)}$.}. The degree of mixing of the spin state is quantified by the purity
\begin{gather}
    \gamma_n \equiv {\rm Tr}\left(\rho_n^2\right)\,.
\end{gather}
In the following, we specifically consider the case where the fermionic state $|\Phi_e\rangle$ is either an Hamiltonian eigenstate (the ground or a low-lying excited state) or a thermal equilibrium state.

The multipartite entanglement of the $n$-spin state is quantified by means of the GME concurrence. For a pure spin state $|\psi_n\rangle$, this is given by the expression~\cite{Ma11b,Huber2010}:
\begin{align}\label{eq:GMEC}
    \mathcal{C}_{\rm GME} (|\psi_n\rangle) 
    \!=\! \min_{\xi\in\Xi} \left\{2\left[ 1\!-\!{\rm Tr}\left( \rho^2_{A_{\xi}} \right) \right]\right\}^{\frac{1}{2}}.
\end{align}
Here $\Xi$ represents the set of all possible bipartitions $\xi$ of the $n$-spin system into the subsystems $A_\xi$ and $B_\xi$, and $\rho_{A_{\xi}}$ is the reduced density matrix of $A_{\xi}$. The GME concurrence vanishes if there is at least one bipartition $A-B$ such that the spin state $|\psi_n\rangle$ can be written in a factorized form, $|\psi_n\rangle=|\Phi_A\rangle\otimes |\Phi_B\rangle$. The theoretical maximum of the GME concurrence for a system of $1/2$ spins is 1, because this value cannot be exceeded whenever the subsystem $A$ coincides with a single spin, being in this case Tr$(\rho_A^2)\ge 1/2$. The GME concurrence and the detection probability are the essential figure of merit, which are used to quantify the GME and the representativeness of the extracted spin state, respectively.

If the spin state $\rho_{n}$ is mixed, as is generally the case, the quantification of entanglement requires a convex-roof extension of the measure defined through Eq. (\ref{eq:GMEC}):
\begin{gather}\label{eq:GMEC_MIXED}
    \mathcal{C}_{\rm GME} (\rho_{n}) = \min_{\{p_i,|\psi_i\rangle\}} \Biggr[\sum_i p_i\, \mathcal{C}_{\rm GME} (|\psi_i\rangle)\Biggr]\,,
\end{gather}
being $\{p_i,|\psi_i\rangle\}$ the set of all decompositions $\sum_i p_i|\psi_i\rangle\langle\psi_i|=\hat\rho_{n}$ of the density matrix. The minimization over all these decompositions represents the most demanding numerical step in the present calculations, and is briefly described in the following Subsection.

The pure spin states $|\psi_n\rangle$ that we consider hereafter are all spin singlets. The mixed spin states $\rho_{n}$, instead, don't have a well defined value of the total spin $S$, but are equally characterized by rotational invariance. This implies that $\rho_{n}$ is a mixture of components corresponding to different values of the total spin. Each of these $S$ components, in turn, consists of a mixture of terms with equal weight, corresponding to all possible values of the total spin projection $M$. In other words, the density operator $\hat\rho_{n}$ is given by the combination of (normalized) projectors on subspaces characterized by given values of the total spin $S$ and of the scalar quantum numbers, collectively denoted by $\alpha$ \cite{Troiani25b}:
\begin{gather}\label{eq:rhos}
\hat\rho_{n} = \sum_S\sum_\alpha p_{\alpha,S} \hat{\mathcal{P}}_{\alpha,S}\,,
\end{gather}
where
\begin{gather}
\hat{\mathcal{P}}_{\alpha,S}=\frac{1}{2S+1}\sum_{M=-S}^S | \alpha,S,M \rangle \langle \alpha,S,M|\,.
\end{gather}
The above expressions include the case where the sum in $\alpha$ runs on an overcomplete basis. In this case, the coefficients $p_{\alpha,S}$ can also take negative values, and the only constraint is given by the normalization condition $\sum_S\sum_\alpha p_{\alpha,S} = 1$.
In the case of a four-spin state ($n=4$), for example, the possible sets of scalar quantum numbers include pairs of partial spin sums $S_{ij}$, each ranging from 0 to 1 and corresponding to the sum of the two spins extracted from the orbitals $(\phi_j,\phi_j)$. More specifically, the partial spin sums corresponding to disjoint pairs (e.g., $S_{12}$ and $S_{34}$) define, together with $S$ and $M$, a complete set of basis states \cite{Tsukerblat}. A sum on an overcomplete basis is one where, for example, $\alpha$ runs not only over different values of $(S_{12},S_{34})$, but also over different values of an incompatible set of quantum numbers, such as $(S_{13},S_{24})$ or $(S_{14},S_{23})$. 

Each of the detection orbitals (modes) $\phi_k$ can be in four different states: unoccupied, doubly occupied, or singly occupied, by a spin-up ($\uparrow$) or spin-down ($\downarrow$) electron. The presence of multipartite entanglement in the spin state $\rho_{n}$ specifically results from the singly-occupied states. The components of the $K$-electron state that include unoccupied detection orbitals do not contribute to the extracted spin state, while doubly-occupied detection orbitals provide mixed contributions to the spin state $\rho_{n}$, which tend to reduce both its purity and the amount of multipartite entanglement \cite{Troiani25b}. This implies a significant difference between the spin entanglement that we are considering and the entanglement between orbital modes, which results from the distribution of the reduced density matrix among all occupation states, and thus tends to increase in the presence of charge fluctuations.

In order to single out the effect of the doubly-occupied orbitals, we consider also the spin states extracted from the component of the projection of the fermionic state on the subspace that includes no double occupations in the detection orbitals:
\begin{gather}\label{eqxxx}
|\Phi_e^*\rangle = \hat{\mathcal{Q}}\, |\Phi_e\rangle \,.   
\end{gather}
Here, the projector $\hat{\mathcal{Q}}$ reads
\begin{gather}
    \hat{\mathcal{Q}} = \prod_{k=1}^n \left(\hat 1_k-\hat n_{k,\uparrow}\,\hat n_{k,\downarrow}\right)\,,
\end{gather}
while $\hat n_{k,\sigma}$ is the number operator corresponding to orbital $\phi_k$ and spin $\sigma=\uparrow,\downarrow$, and $\hat 1_k$ is the identity operator in the four-dimensional space of the $k$-th mode. In the following, we refer with $\Gamma^{(n,*)}_{\bf\sigma,\bf\sigma'}$ to the $n$BRDM obtained by replacing $\Phi_e$ with $\Phi_e^*$ in Eq. (\ref{eqn:nBRDM}), with $p_n^*=w_n^*={\rm Tr}\big[\Gamma^{(n,*)}_{\bf\sigma,\bf\sigma'}\big]$ to the detection probability, and with $\rho_n^*$ to the normalized spin state. The projector $\hat{\mathcal{Q}}$ nullifies a component of the fermionic state $|\Phi_e\rangle$ that can contribute to the $n$BRDM, and therefore reduces the trace of the $n$BRDM, i.e. the unnormalized detection probability: $w^*_n\le w_n$.

\subsection{Numerical approach} \label{sec:numerics}

The numerical evaluation of the convex-roof entanglement measure can be implemented through a conjugate gradient algorithm in the space of unitary matrices \cite{Audenaert01,rothlisberger2009numerical}. For the reader's convenience, we report in the following a brief outline of this approach. A more detailed discussion can be found in the cited papers.

In order to perform the minimization [Eq. ~\eqref{eq:GMEC_MIXED}], the decompositions of the spin state $\rho_{n}$ have to be parametrized. This is achieved through the Schrödinger-Hughston-Jozsa-Wootters theorem \cite{Hughston93a}, which establishes a biunivocal correspondence between the decompositions of $\rho_{n}$ and the set of $q\times r$ unitary matrices $U$ (formed by $r$ orthonormal columns, such that $U^\dagger U=I$), being $r$ the rank of $\rho_{n}$ and $q\ge r$ the number of terms in the decomposition. The correspondence is defined by the equation:
\begin{gather}
    |\psi_i\rangle = \frac{1}{\sqrt{p_i}}\sum_{j=1}^r U_{ij} \sqrt{\lambda_j}\,|\chi_j\rangle
\end{gather}
where $\lambda_i$ and $|\chi_i\rangle$ are the nonzero eigenvalues and the corresponding eigenstates of $\rho_{n}$, respectively, and
\begin{gather}
    \hat\rho_{n} = \sum_{j=1}^r \lambda_j |\chi_j\rangle\langle\chi_j| = \sum_{i=1}^q p_i|\psi_i\rangle\langle\psi_i| \,.
\end{gather}
Here, both the probabilities $p_i$ and $\lambda_j$ normalize to 1, and the states $|\psi_i\rangle$, unlike the $|\chi_j\rangle$, are not mutually orthogonal.
The minimization can thus be performed in the space of the unitary matrices $U$. 

This minimization is numerically implemented through a conjugate-gradient method adapted to the unitary group of the square ($q \times q$) matrices, which represent an overparametrized counterpart of the original $q \times r$ ones (where the last $q-r$ columns do not affect the expression of the density operator). The actual search is then implemented in the space of the skew-Hermitian matrices $X$ through the expression $U=\exp(tX)$; here $t$ is the real parameter over which the local minimization is performed, while $X$ defines at each step the search direction. 

The lowest value of the GME concurrence obtained for a given value of $q$, $\mathcal{C}_{\rm GME}^{(q)}$, represents an upper bound for the GME concurrence. In the minimization procedure [Eq. (\ref{eq:GMEC_MIXED})], one should in principle consider decompositions including arbitrary values of $q$. In practice, the size of the matrices, corresponding to the number of states in the decomposition of $\rho_n$, is increased until $\mathcal{C}_{\rm GME}^{(q)}$ saturates; the lowest value obtained is then identified with $\mathcal{C}_{\rm GME}$. In the following calculations, the rank $r$ of $\rho_s$ is of the order of a few tens for the ground states, it is sufficient to set $q$ to $r+5$ in order to obtain convergence ($\mathcal{C}_{\rm GME}^{(q)}-\mathcal{C}_{\rm GME}^{(q+1)}\lesssim 10^{-3}$), and the number of random initial points that initialize the search of the minimum in the space of the skew-Hermitian matrices is of the order of ten. 

\section{Model systems and results\label{sec:results}}

\begin{figure}[tbp]
\includegraphics[width=0.47\textwidth]{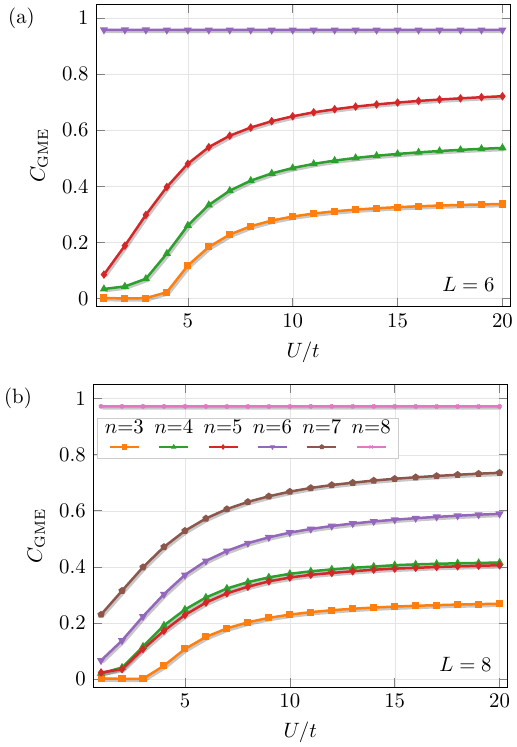}
\caption{Genuine multipartite entanglement concurrence of the spin state extracted from the ground state of the Hubbard ring at half filling ($K=L$), with $V=0$, and for different values of the spin number $n$. The two panels correspond to: (a) $L=6$ and (b) $L=8$ sites. The same color code for the spin number $n$ is adopted in the two panels.}
\label{fig01}
\end{figure} 

In the following, we investigate the GME in the standard and in the extended Hubbard models with a ring geometry. These are defined by the Hamiltonian
\begin{gather}
\hat{H}_e = \sum_{k=1}^L \,\bigr\{U\,\hat{n}_{k,\uparrow}\,\hat{n}_{k,\downarrow} + V\,\hat{n}_{k}\,\hat{n}_{k+1} \nonumber\\ \,\sum_{\sigma=\uparrow,\downarrow} \bigr[\epsilon\,\hat{n}_{k,\sigma} - t (\hat{c}_{k,\sigma}^\dagger\,\hat{c}_{k+1,\sigma} + \hat{c}_{k+1,\sigma}^\dagger\,\hat{c}_{k,\sigma})\bigr]\bigr\}\,,
\end{gather}
with zero and nonzero values of $V$, respectively.
In the expression of the Hamiltonian, $\hat{n}_{k}=\hat{n}_{k,\uparrow}+\hat{n}_{k,\downarrow}$ and
$\hat{n}_{k,\sigma}=\hat{c}^\dagger_{k,\sigma}\,\hat{c}_{k,\sigma}$ is the number operator corresponding to the $k$-th site within the ring ($1\le k\le L$) and to the spin state $\sigma=\uparrow,\downarrow$. The symbols $t$, $U$, and $V$ denote the hopping, on-site and nearest-neighbor interaction parameters, respectively. In view of the periodic boundary conditions, the site number $L+1$ coincides with the site number $1$. 

For the Hubbard model the detection orbital $\phi_k$ mentioned in the previous Section coincides with the one localized on the $k$-th site of the ring. In all the cases considered hereafter, the spins are extracted from neighboring sites, and specifically from the $n$ consecutive orbitals $\phi_1,\phi_2,\dots,\phi_n$. Due to the translation invariance of the system, the resulting $n$BRDM is identical to the ones obtained with any other set of consecutive orbitals.

\subsection{\texorpdfstring{Standard Hubbard model \\ at half filling}{Standard Hubbard model at half filling}}

\begin{figure}[tbp]
\centering
\includegraphics[width=0.47\textwidth]{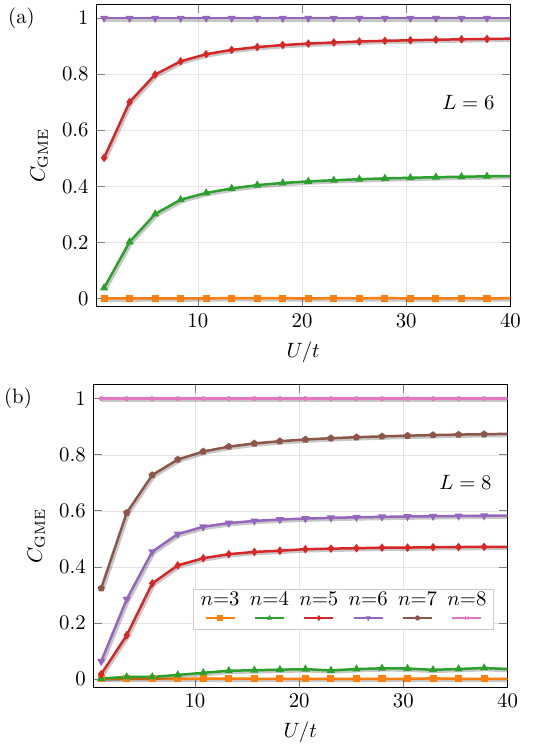}
\caption{Genuine multipartite entanglement concurrence of the spin state extracted from the first excited singlet state of the Hubbard ring at half filling ($K=L$), with $V=0$, and for different values of the spin number $n$. The two panels correspond to: (a) $L=6$ and (b) $L=8$ sites. The same color code for the spin number $n$ is adopted in the two panels.}
\label{fig02}
\end{figure} 

We start by considering the entanglement properties of the ground state $|\Phi_{e,0}\rangle$ as a function of $U/t$, for $V=0$ and $K=L$. As previously shown~\cite{Troiani25a}, when the number of extracted particles coincides with the number of electrons in the system ($n=K$), the GME concurrence approaches its theoretical maximum, which coincides with that of the Heisenberg ring, irrespective of the value of $U/t$. Here we find that, in all the other cases ($n<K$), the GME concurrence vanishes in the non-interacting limit ($U\ll t$) and increases monotonically with $U/t$ (Fig. \ref{fig01}). This trend qualitatively corresponds to that of the two-spin concurrence, but is opposite with respect to the one displayed by the von Neumann entropy of the site, which displays a pronounced maximum for $U=0$ \cite{Gu04a}. Such a discrepancy confirms the striking differences between spin and orbital-mode entanglement.

The behavior of the GME can be interpreted as follows. In the non-interacting limit ($t\gg U$), the spin states $\rho_{n}$ (for any $n<K$) extracted from the ground state of the Hubbard ring are maximally mixed and thus fully separable. In the opposite limit ($t \ll U$), the contribution of the doubly occupied states becomes negligible. Therefore, the spin state $\rho_{n}$ with $n=K$ corresponds to the (multipartite entangled \cite{Siloi14a}) ground state of the Heisenberg model. Besides, those with $n<K$ tend to the reduced states obtained from the Heisenberg ground state, which are also, to some extent, multipartite entangled. The monotonic increase of the GME concurrence with $U/t$ can be regarded as a consequence of the increased purity ($\gamma_n$) of the spin states $\rho_{n}$, which in turn results from the reduced presence of ``ionic contributions'' (configurations with doubly-occupied sites) in the ground state $|\Phi_{e,0}\rangle$. This interpretation is confirmed by the comparison with the GME concurrences of the spin state $\rho^*_n$, obtained after projecting out the components of the $K$-electron state with double occupation in the detection orbitals [Eq. (\ref{eqxxx})]. In this case, in fact, the GME concurrence is essentially independent on $U/t$ for all the values of $n$. The difference with respect to the complete fermionic state is negligible in the limit $U\gg t$, where the double occupations in $|\Phi_e\rangle$ are suppressed by the Coulomb interactions, but becomes significant in the weakly-interacting regime. An indirect confirmation of the role played by the double occupations is also provided by the behavior of $\mathcal{C}_{GME}(\rho_n)$ for $n=K$ (Fig. \ref{fig01}): here the ionic configurations do not contribute to the expression of the $n$BRDM, and a change in their weight does not affect the value of the GME concurrence \cite{footnote}.

\begin{figure}[t]
\centering
\includegraphics[width=0.47\textwidth]{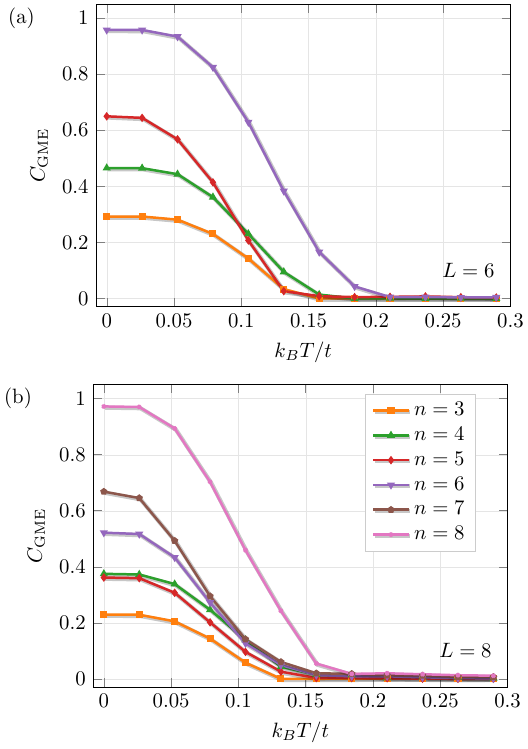}
\caption{Genuine multipartite entanglement concurrence of the spin state extracted from the thermal equilibrium state $\tau$ of the Hubbard ring. The concurrence is plotted as a function of the temperature $T$, normalized to $t/k_B$ (with $k_B$ the Boltzmann factor). The concurrence is computed for different values of the spin number $n$, at half filling ($K=L$), with $U/t =10$ and $V=0$. The two panels correspond to: (a) $L=6$ and (b) $L=8$ sites in the ring. The same color code for the spin number $n$ is adopted in the two panels.}
\label{fig03}
\end{figure} 

The GME concurrence systematically increases, for each given value of $U/t$, with the number of spins $n$. This can also be correlated with the purity of the spin state: the larger the number of extracted electrons $n$, the higher the value of $\gamma_{n}$, the more effectively the detection process selects a spin-entangled component from the fermionic state $|\Phi_{e,0}\rangle$. The dependence of the purity on $n$ displays in fact a sawtooth, rather than a monotonic behavior, because $\rho_n$ states with even values of $n$ tend to have a higher values of $\gamma_n$ than those with odd $n$.

In the energy spectrum of the considered Hubbard rings, the singlet ground state is followed by a spin triplet and by an excited singlet. This presents a clear ``redistribution'' of entanglement with respect to the ground state (Fig. \ref{fig02}). In fact, on the one hand, the full particle extraction ($n=K$) leads to a maximal GME concurrence ($\mathcal{C}_{GME}=1$), and the values of  $\mathcal{C}_{\rm GME}$ are higher than those obtained for the ground state when $n \gtrsim K/2$.
On the other hand, the values of the GME concurrence are lower than the corresponding ones obtained for the ground state when $n \lesssim K/2$. The negative correlation between these two forms of entanglement (involving either small or large numbers of spins) can be regarded as a manifestation of the monogamy of entanglement \cite{Coffman00a}.

We conclude the analysis of this model by computing the GME extracted from thermal states, i.e. from fermionic states 
\begin{gather}
\hat\tau = \frac{1}{Z}\, e^{- \hat H_e/k_BT} = \frac{\sum_{k=0}^\infty |\Phi_{e,k}\rangle\langle\Phi_{e,k}| \,e^{-E_k/k_BT}}{\sum_{l=0}^\infty e^{-E_l/k_BT}} \,,
\end{gather} 
where $E_k$ and $|\Phi_{e,k}\rangle$ are the eigenvalues and eigenstates of $ H_e$, respectively. 

We specifically consider the Hubbard ring in the Mott-insulating phase ($U/t=10$), because this is the region where the detected spin state displays a large amount of GME. In this regime, the energy gap $\Delta$ between the ground and first excited (triplet) states corresponds to a spin --- rather than a charge --- excitation, with a characteristic energy scale $4t^2/U$ \footnote{The energy gaps in the Heisenberg rings are 0.685$\,J$ and 0.523$\,J$ for $L=6$ and $L=8$, respectively.}.  
The GME concurrences of all the spin states $\rho_{n}$ decrease monotonically with the temperature. More specifically, $\mathcal{C}_{\rm GME}$ remains constant for $k_BT\ll\Delta$, thus displaying a plateau protected by the energy gap, and drops to zero as $k_BT$ approaches $\Delta$ (Fig. \ref{fig03}), as a consequence of the mixing between the singlet and the triplet states. Being the value of $\Delta$ larger for $L=6$ than for $L=8$ (and, in general, a decreasing function of $L$ in rings with even site numbers), the decay of GME takes place at higher temperatures. We note that, also at finite temperatures, the value of the GME concurrence increases with the number of extracted spins $n$, as observed for the system ground state. Finally, as expected in view of the highly entangled character of the ground state and of the mixed character of the triplet contribution, the system lacks the anomalous non-monotonic thermal enhancement seen in highly frustrated or spin-orbit-coupled systems.

\subsection{Effect of the positive doping (\texorpdfstring{$K<L$}{K<L})}

\begin{figure}[b]
\centering
\includegraphics[width=0.47\textwidth]{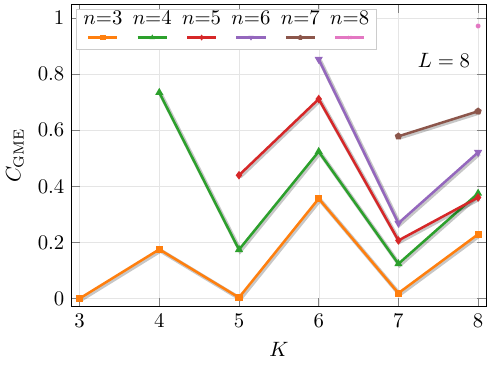}
\caption{Spin entanglement extracted from the ground state multiplet of the Hubbard ring, as a function of $K>L$ (positive doping) and for different values of the spin number $n$, with $U/t=10$, $L=8$, and $V=0$.} \label{fig04}
\end{figure} 

\begin{table}[tbp]
\caption{Ground state of the Hubbard ring, $n$BRDMs, and extracted spin states for $L=8$, $U/t=10$, $V=0$, as a function of the number of electrons $K$ (see also Fig.~\ref{fig04}). For each state the table reports: the total spin $S$ of the ground state; its degeneracy $g$; the unnormalized detection probability $\omega_n$, corresponding to the trace of the $n$BRDM; the purity $\gamma_{n}$ of the normalized spin state. For $n=2$ (nearest neighbours) $\mathcal{C}_{\mathrm{GME}}$ coincides with the Wootters concurrence. }
\vspace{0.1cm}
\label{tab:doped_K}
\begin{ruledtabular}
\begin{tabular}{ccccccc}
$K$ & $S$ & $g$ & $n$ & $\omega_n$ & $\gamma_{n}$ & $\mathcal{C}_{\mathrm{GME}}$ \\
\hline
3 & 3/2 & 4 & 2 & $4.96\times10^{-2}$ & 0.333 & 0.000 \\
  &     &   & 3 & $1.34\times10^{-3}$ & 0.250 & 0.000 \\
\hline
4 & 1   & 3 & 2 & 0.149               & 0.394 & 0.157 \\
  &     &   & 3 & $2.27\times10^{-2}$ & 0.344 & 0.175 \\
  &     &   & 4 & $1.09\times10^{-3}$ & 0.333 & 0.735 \\
\hline
5 & 1/2 & 4 & 2 & 0.298               & 0.482 & 0.334 \\
  &     &   & 3 & 0.102               & 0.352 & 0.003 \\
  &     &   & 4 & $2.18\times10^{-2}$ & 0.306 & 0.174 \\
  &     &   & 5 & $2.09\times10^{-3}$ & 0.264 & 0.440 \\
\hline
6 & 0   & 1 & 2 & 0.494               & 0.553 & 0.454 \\
  &     &   & 3 & 0.286               & 0.445 & 0.356 \\
  &     &   & 4 & 0.138               & 0.570 & 0.524 \\
  &     &   & 5 & $5.02\times10^{-2}$ & 0.493 & 0.712 \\
  &     &   & 6 & $1.08\times10^{-2}$ & 1.000 & 0.853 \\
\hline
7 & 1/2 & 4 & 2 & 0.728               & 0.467 & 0.306 \\
  &     &   & 3 & 0.590               & 0.359 & 0.019 \\
  &     &   & 4 & 0.459               & 0.319 & 0.124 \\
  &     &   & 5 & 0.335               & 0.260 & 0.207 \\
  &     &   & 6 & 0.217               & 0.225 & 0.269 \\
  &     &   & 7 & 0.106               & 0.252 & 0.579 \\
\hline
8 & 0   & 1 & 2 & 0.976               & 0.504 & 0.373 \\
  &     &   & 3 & 0.950               & 0.415 & 0.229 \\
  &     &   & 4 & 0.926               & 0.457 & 0.375 \\
  &     &   & 5 & 0.902               & 0.414 & 0.360 \\
  &     &   & 6 & 0.879               & 0.504 & 0.521 \\
  &     &   & 7 & 0.855               & 0.484 & 0.668 \\
  &     &   & 8 & 0.811               & 1.000 & 0.972
\end{tabular}
\end{ruledtabular}
\end{table}

Before considering the effect on the GME of coupling between nearest neighbors, we investigate that of positive doping ($K<L$). While the ground state of the Hubbard model at half filling corresponds to a (nondegenerate) spin singlet, this is no longer the case for $K<L$. The spin extraction is thus performed not from a pure fermionic state $|\Phi_e\rangle$, but from a mixed state $\tau$, where the $2S+1$ components of the multiplet contribute with an equal weight. Additional, symmetry-related $\kappa$-fold degeneracies are also possible, giving rise to an overall degeneracy $g=\kappa(2S+1)$.
Equation (\ref{eqn:nBRDM}) thus generalizes to: 
\begin{gather}\label{eqn:nBRDMmixed}
    \Gamma^{(n)}_{{\bf\sigma},{\bf\sigma}'} \!=\frac{1}{g}\!\sum_{a=1}^\kappa\sum_{M=-S}^S \langle \Phi^{\alpha,M}_{e,0} | \,\hat{c}_{n,\sigma_n'}^\dagger \dots \hat{c}_{n,\sigma_n}\, |\Phi_{e,0}^{\alpha,M}\rangle\;,
\end{gather}
being $|\Phi_{e,0}^{\alpha,M}\rangle$ the fermionic ground state corresponding to total spin projection $M$ and $S$-multiplet index $\alpha$. This corresponds to the thermal equilibrium state in the zero-temperature limit, i.e. when $k_BT$ is much smaller than the energy gap between the ground and first-excited multiplets. In general, the mixed character of the fermionic state tends to reduce the purity of the extracted spin state $\rho_{n}$, and thus the amount of GME. However, the relation between the GME and the purity of the fermionic state, given by $1/g$, is not straightforward, nor is that between $g$ and the purity of the spin state $\rho_n$ \footnote{Orthogonal and equally weighted fermionic states can give rise to spin states ($n$BRDMs) that are neither orthogonal, nor equally weighted. Therefore, the purity of the spin states $\rho_n$ cannot be directly deduced from the value of $g$.}.

As a representative example, we consider the ring with $L=8$ in the Mott-insulating phase ($U/t=10$). The values of $S$ for the ground states corresponding to different values of $K$ are reported in Table \ref{tab:doped_K}, along with other parameters that characterize the $n$BRDM and the extracted spin state $\rho_{n}$. The dependence of the GME concurrence on the number of electrons $K$ displays a clear sawtooth behavior for each spin number $n\le K$ (Fig. \ref{fig04}). In fact, for any given $n$, the values of $\mathcal{C}_{\rm GME}$ obtained for system with odd values of $K$ are systematically lower than those obtained for even values. This applies also to the system with four electrons, whose ground state has a larger spin ($S=1$) than the five-electron system, but a higher degeneracy $g$. 

The presence of holes in the Hubbard ring induces charge fluctuations that, at half filling, are frozen by the strong Coulomb interactions ($U/t=10$). These charge fluctuations tend to increase the amount of entanglement between the sites of the ring, but have a less univocal effect on the extracted spin entanglement that we are currently investigating. In fact, on the one hand, the presence of doubly occupied sites tends to decrease the purity of the spin states, and thus the amount of entanglement in the $\rho_{n}$. On the other hand, the charge fluctuations that result in unoccupied detection sites $l\le n$ are irrelevant, because they do not contribute to the $n$BRDMs, and thus do not affect the extracted spin states $\rho_{n}$ \cite{footnote}. 

In any case, doping does not monotonically reduce the GME entanglement of the extracted spin state, which is maximized for all values of $n\le K$ by the presence of two holes ($K=6$). On the other hand, the positive doping systematically reduces the detection probability: this decreases strongly with $n$ for $K<L$, and much more weakly in the undoped system ($K=L$).
We finally note that, also in this case, there are clear correlations between the dependence on $n$ and $K$ of the GME concurrence and that of the purity (Table \ref{tab:doped_K}).

\subsection{\texorpdfstring{Extended Hubbard model \\ at half filling}{Extended Hubbard model at half filling}}

The extended Hubbard model is known to display quantum phase transitions, which result from the competition between on-site and inter-site repulsion, and finite-size precursors of such transitions in systems formed by a limited number of sites \cite{Gu04a}. If the on-site repulsion prevails, the system is in the Mott-insulator phase, antiferromagnetic spin correlations show up, and the gap in the infinite-size limit is related to the charge degrees of freedom (spin-density waves). If the inter-site repulsion dominates, charge fluctuations are maximal (alternation of unoccupied and doubly-occupied sites), and a gap opens both for the charge and the spin degrees of freedom (charge-density waves). In the strong-coupling regime ($U,V\gg t$), spin- and charge-density wave regimes are obtained for $V\lesssim U/2$ and $V\gtrsim U/2$, respectively. In the following, we focus on the transition between these two regimes and on its impact on the GME of the extracted spin states.

\begin{figure}[b]
\centering
\includegraphics[width=0.47\textwidth]{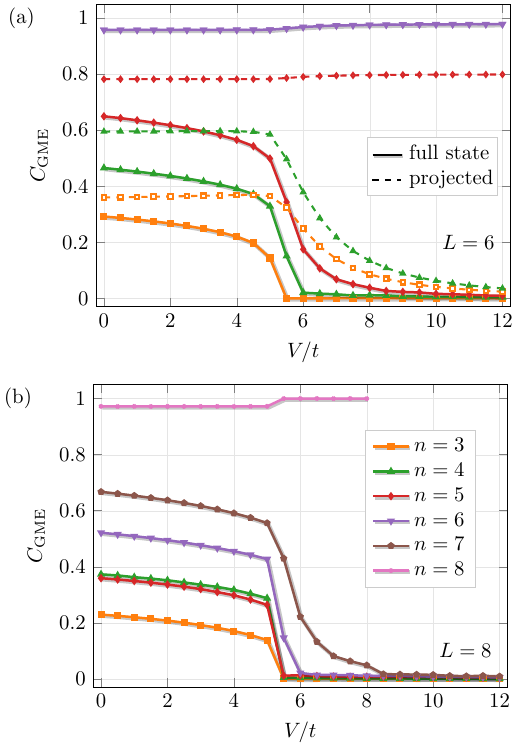}
\caption{Genuine multipartite entanglement concurrence extracted from the ground state of the Hubbard ring as a function of $V/t$, for different values of the spin number $n$, with $K=L$ and $U/t=10$. The two panels correspond to: (a) $L=6$ and (b) $L=8$ sites in the ring. The same color code for the spin number $n$ is adopted in the two panels.}
\label{fig05}
\end{figure}

The transition spin- and charge-density wave regimes is fully captured by the GME concurrence, which, precisely in the region $V\approx U/2$, undergoes an abrupt reduction for $n<K$, and displays a kink for $n=K$ (Fig. \ref{fig05}). This transition is observed both for the $L=6$ and for the $L=8$ rings, and is sharper in the latter case. In all the considered range of values, the GME concurrence increases with the size of the spin system ($n$), as in the case $V=0$ discussed above. 

\begin{table}[t]
\caption{Unnormalized detection probability $w_n$, corresponding to the trace of the $n$BRDMs, obtained from the ground state of the extended Hubbard model, for $U/t=10$ and $K=L$.}
\label{tab:weights}
\begin{ruledtabular}
\begin{tabular}{|c|c|c|c|}
$V/t$ & $L=6$, $n=6$ & $L=8$, $n=6$ & $L=8$, $n=8$ \\
\hline
4   & $0.668$                & $0.739$                & $0.616$             \\
5   & $0.405$                & $0.654$                & $0.514$             \\
6   & $5.16\times10^{-3}$    & $1.11\times10^{-3}$    & $2.51\times10^{-4}$ \\
7   & $3.32\times10^{-4}$    & $1.05\times10^{-4}$    & $7.97\times10^{-6}$ \\
8   & $5.33\times10^{-5}$    & $2.09\times10^{-5}$    & $7.46\times10^{-7}$ \\
9   & $1.33\times10^{-5}$    & $5.69\times10^{-6}$    & $\le 10^{-7}$       \\
10  & $4.32\times10^{-6}$    & $2.05\times10^{-6}$    & $\le 10^{-7}$       \\
\end{tabular}
\end{ruledtabular}
\end{table}

The reduction of the GME concurrence results from the alternation of unoccupied and doubly-occupied orbitals that characterizes the ground state in the charge density wave regime. Here, not only the extracted spin state $\rho_{n}$ displays a negligible amount of GME, but the probability of extracting $n$ spins from $n$ consecutive sites is negligible. This is clearly witnessed by the sudden drop in the values of $w_n$ as $V$ becomes larger than $U/2$ (Table \ref{tab:weights}). 

The comparison with the GME of the spin state $\rho^*_n$, obtained after removing the double occupations from the fermionic state [Eq. (\ref{eqxxx})], sheds further light on the transition that takes place at $V\approx U/2$. Here, in fact, the states with low spin numbers display a less sudden drop in $\mathcal{C}_{\rm GME}$ with respect to $\rho_n$, and are in general more entangled. Besides, high spin numbers ($n=5,6$) undergo an increase in the GME concurrence [Fig. \ref{fig05}(a), dotted lines], even if this region corresponds to very low values of the detection probability, and is thus less significant (Table \ref{tab:weights}). In any case, the comparison between the GME obtained with ($\rho_n^*$) and without ($\rho_n$) the projection shows that the effect of the phase change on the spin state cannot be entirely reduced to the impurity induced by double occupations.

\subsection{A molecular ring: the case of benzene\label{subsec:benzene}}

\begin{table}[b]
    \centering
    \begin{tabular}{|c|c|c|c|c|}
\hline
$\mathcal{C}_{\rm GME} (\rho_{n})$ & \multirow{2}{3em}{$n=3$} & \multirow{2}{3em}{$\ n=4$} & \multirow{2}{3em}{$\ n=5$} & \multirow{2}{3em}{$\ n=6$} \\
$[{\rm Tr}(\rho^2_s)]$ & & & & \\
\hline
Ground state & $0.000314$ & $0.0587$ & $0.253$ & $0.958$ \\
             & $(0.318)$ & $(0.330)$ & $(0.348)$ & $(1.00)$ \\
Excited state & $0.000458$ & $0.135$ & $0.614$ & $1.00$ \\
             & $(0.235)$ & $(0.256)$ & $(0.367)$ & $(1.00)$ \\
\hline
    \end{tabular}
    \caption{Genuine multipartite entanglement concurrence obtained for the ground and first excited singlet states of the benzene molecule. Different columns correspond to different numbers of spins ($n$), extracted from orthogonal atomic orbitals localized on consecutive C atoms. The numbers in parentheses are the purities of the extracted spin states. }
    \label{tab:1}
\end{table}
 
To complement the analysis of model Hamiltonians, we consider the benzene molecule (C$_6$H$_6$). Reference~\cite{Troiani25a} demonstrated that near-maximal spin GME can be extracted from its ground state. Related studies characterized pairwise entanglement between localized orbitals~\cite{Ding23a} and near-maximal GME among the six localized $\pi$ orbital modes~\cite{Ding26a}. Here we extend the extracted-spin analysis to smaller subsets of electrons, whose spin states are mixed, and examine how closely their entanglement is reproduced by the Hubbard model. In particular, we consider the ground and first excited singlet states derived from Complete Active Space Self Consistent (CASSCF) calculations. The detection orbitals $\phi_i$ that define the $n$BRDM coincide with the orthogonal atomic orbitals that maximize the overlap with the $2p_z$ orbitals of the six C atoms (further details on the computational approach can be found in the Supplemental Material of Ref. \onlinecite{Troiani25a}).  

If the number of electrons extracted coincides with the number of electrons in the active space ($n=K=6$), the spin-entanglement properties of the ground state and the first excited singlets are identical to those of the Hubbard model, up to numerical accuracy \cite{Troiani25a}. The near-maximal six-spin concurrence in Table~\ref{tab:1} characterizes the GME of the spin state obtained by selecting one electron per detection orbital. The correspondence with the Hubbard model is no longer perfect if smaller numbers of extracted spins are considered. However, also for $n\le K$, the results for the benzene molecule (Table \ref{tab:1}) approximately coincide with those obtained for the six-site Hubbard ring with $U\approx 2t$, both for the ground and for the first excited singlet states. This suggests that the entanglement properties of the extracted spin states are largely independent on the details of the system Hamiltonian. 

  \section{Conclusions\label{sec:conclusions}}

In conclusion, we have quantified the genuine multipartite entanglement extracted from the lowest eigenstates and from the thermal state of Hubbard rings. This analysis deals with mixed spin states, thus going beyond the pure states considered in previous studies~\cite{Troiani25a,Troiani25b}.

We have shown that, in Hubbard rings at half filling, the on-site repulsion enhances the purity of the spin states and their GME by suppressing charge fluctuations, and that both these quantities grow with the number $n$ of extracted spins. The suppression of charge fluctuations also enhances the detection probability, making this regime ideal for the extraction of multipartite-entangled spin states. In the strongly interacting regime, the GME is robust at temperatures below the singlet--triplet gap, while the detection probability decreases only weakly with $n$.

Doping can even enhance the GME, in spite of the induced charge fluctuations, but at the cost of a detection probability that decreases rapidly with $n$. Conversely, the repulsion between neighboring sites suppresses both quantities, by inducing an alternation of unoccupied and doubly occupied sites, and thus sharply signals the transition between spin- and charge-density-wave regimes. Finally, the GME extracted from the benzene molecule is approximately reproduced by that of a six-site Hubbard ring with $U\approx 2t$, also for $n<K$.

Overall, the strongly interacting regime at half filling emerges as the one that best combines a high GME with a high detection probability.

\acknowledgments

We acknowledge financial support from: the European Union’s Horizon Europe research and innovation programme under the Marie Skłodowska-Curie grant agreement number 101120240; the Ministero dell'Università e della Ricerca (MUR) under the Project PRIN 2022 No. 2022W9W423. 

Digital tools incorporating large language models developed by Anthropic and OpenAI were used by the authors to assist with various aspects of this work, including manuscript editing and coding activities, particularly the development of test cases. The authors retain full responsibility for the scientific content, including the design of the study, the analysis and interpretation of the results, and the conclusions drawn.


\begin{thebibliography}{51}
\bibitem [{foo()}]{footnote}%
  \BibitemOpen
  \href@noop {} {}\bibinfo {note} {At half filling, the double occupation of a site $k$ implies the presence of an unoccupied site $l$. Thus, the component of the fermionic state $|\Phi_e\rangle$ would be nullified by the action of the annihilation operator $\hat c_l$ in the expression of the $n$BRDM [Eq. \ref{eqn:nBRDM}], for $n=K$. This mechanism does not apply to the cases $n<K$, if the unoccupied site is not involved in the electron detection ($l>n$).}\BibitemShut {Stop}%
\end{thebibliography}

%

\end{document}